\documentclass[12pt,a4paper,final]{iopart}
\usepackage{iopams}  
\usepackage{graphicx}
\usepackage{dcolumn}  
\usepackage{bm}       
\usepackage{amssymb} 
\usepackage{braket}
\usepackage{mathcomp}

\usepackage{bbold}
\expandafter\let\csname equation*\endcsname\relax
\expandafter\let\csname endequation*\endcsname\relax 
\usepackage{amsmath}
\usepackage{physics}

\begin{document}

\title[Preparing an article for IOP journals in  \LaTeXe]{The non-orientable spacetime of the eternal black hole}
\vspace{3pc}

\author{Ovidiu Racorean}
\address{General Direction of Information Technology}
\address{Banul Antonache str. 52-60, sc.C, ap.19, Bucharest, Romania}
\ead{ovidiu.racorean@mfinante.gov.ro}
\vspace{3pc}

\begin{abstract}
\vspace{1pc}

We derive the metric of the BTZ black hole for the special case of interchanging the characteristics of space and time coordinates. We maximally extend the geometry of this interior solution and show that the line element is similar to its exterior solution counterpart. Thus, we assign the thermofield double state dual to this interior geometric construction. As a result, we have two independent thermofield double states corresponding to the BTZ metric: one dual to the exterior solution and one dual to the interior solution. In this scenario, considering both thermofield double states as dual to the full BTZ black hole, we evaluate the partition function of the bulk. The partition function represents a non-orientable spacetime. Additionally, we derive a thermofield double-like state that connects in the gravity dual the regions of spacetime with opposite orientations of space and time.

\end{abstract}

%\pacs{04.70.Dy, 03.67.Bg, 42.50.Ex, 95.30.Gv}
%\vspace{7pc}
%\noindent{}      \hspace{7pc}          Essay written for the Gravity Research Foundation 
%\vspace{1pc}

%\noindent{}       \hspace{10pc}              2023 Awards for Essays on Gravitation

%\noindent{}  \hspace{15pc}  March 31, 2023
%\submitto{ Essay written for the Gravity Research Foundation 2017 Awards for Essays on Gravitation}
\maketitle

\section{Introduction}

The concept of interchange the characteristics of space and time coordinates has deep roots that extend across various fields of physics research. As such, in general relativity, the switching of space and time roles occurs behind the event horizon of Schwarzschild black holes' solution. The space and time Schwarzschild coordinates $(t, r)$ interchange their roles within the gravitationally trapped region, $0 < r < 2M$, \cite{Breh}, \cite{Doran}. Moreover, the Schwarzschild metric resulting from the interchange of space and time roles was proposed in the cosmological context in \cite{Doran}, \cite{cule}, \cite{chen}.

However, within the framework of quantum mechanics theory, switching the roles of space and time makes the spatial and temporal correlations equivalent. Specifically, spatial correlations, i.e. the quantum entanglement of two subsystems separated by a spatial distance, are equivalent to one quantum system measured at two different time instances separated by a temporal distance \cite{zhao}, \cite{Zhang}, \cite{Leifer}, \cite{full}.

Since it is related to our work in this paper, we should emphasize here the interchange of the characteristics of space and time coordinates in the context of string theory that distinguishes between the sectors of open and closed strings \cite{blum}, \cite{angel}.

In this paper, we derive the BTZ line element in the context of interchanging the space and time roles as seen by an observer situated behind the event horizon, i.e., the interior BTZ solution. Although we found some related work in \cite{berr}, \cite{ryu}, \cite{wei}, \cite{hao}, our derivation of the BTZ interior solution differs significantly. To deepen the understanding of the physical implications drawn from this model, we derive the maximal extension of the interior BTZ spacetime. We show that the line element is similar to its exterior solution counterpart. Furthermore, we discuss the implications of the interchange of space and time roles on the internal maximal extension line element.

To establish a connection with the last sections of the paper, we construct the Euclidean metric of the interior solution. Based on similarities with the exterior solution \cite{mal}, \cite{suss}, we demonstrate that the interior BTZ spacetime is dual to a thermofield double state. Specifically, we derive the thermofield double state in the customary manner from the Euclidean path integral \cite{har}. In this scenario, the entanglement between boundary CFTs connects in the gravity dual the asymptotic region $II$ of the black hole spacetime to the region $IV$, regions that are usually associated to black hole and white hole regions. 

Now, to sum up, we have found that the geometry of spacetime behind the BTZ horizon is dual to the thermofield double state. On the other hand, the typical thermofield double state connects region $I$ and region $III$. Therefore, to cover the entire spacetime of the BTZ black hole, we should account for both thermofield double states. In the full BTZ black hole geometry scenario, we evaluate the partition function of the bulk in the Euclidean path integral formalism. The partition function represents a non-orientable spacetime of the bulk, specifically the Klein bottle geometry in our case, as identified in \cite{berr}, \cite{ryu}, \cite{wei}, \cite{hao}. From this perspective, our work can be related to the study of topological invariants in many-body topological phases of matter protected by orientation-reversing symmetry \cite{shi}, \cite{shap}, \cite{cho}, \cite{poll}. We can find connections between our work and geons in AdS/CFT if we consider a single-sided, asymptotically AdS3 black hole \cite{mona}.

We conclude our work by deriving the thermofield double-like state that connects the regions of spacetime with opposite orientations of space and time in the gravity dual.

\section{The interior solution of the BTZ black hole}

We would like to derive the line element of 2+1 dimensional spacetime as seen by an observer situated behind the event horizon. In such spacetime we should assume the interchanging of space and time roles. Thus, the time coordinate is now r, while the radial coordinate is t and the angular coordinate $\phi$ is $0<\phi<2\pi$ such that we start by considering a general radially symmetric line element:

\begin{equation}
ds^2=-B^2(t)dt^2+A^2(t)dr^2+F^2(t)d\phi^2.
\end{equation}

Taking into account the spherical symmetry and that $t$ is the radial coordinate we rewrite the metric as:

\begin{equation}
ds^2=-B^2(t)dt^2+A^2(t)dr^2+t^2d\phi^2,
\end{equation}

and solve for the functions $B(t)$ and $A(t)$, respectively.

Indeed, after we determine the curvature two-form and solve equation for the empty space with the cosmological constant $\Lambda=-\frac{1}{\ell^2}$ (see  APPENDIX A) we recover functions $A$ and $B$ as:

\begin{equation}
B^2(t) =\frac{1}{M-\frac{t^2}{\ell^2}} .
\end{equation}

and
\begin{equation}
A^{2}(t)=M-\frac{t^2}{\ell^2}.
\end{equation}

where $M$ is a constant of integration.

With these results at hand we can write the metric of the 2+1 black hole as:

\begin{equation}
ds^2=-\frac{\ell^2}{t^{2}_{h}-t^2}dt^2+\frac{t^{2}_{h}-t^2}{\ell^2}dr^2+t^2d\phi^2
\end{equation}

Here we noted $M\ell^2=t^{2}_{h}$ such that $t \rightarrow t_h$ is the event horizon, whereas we still have a singularity at $t \rightarrow 0$ in the region with $t<t_h$ which is the classical exterior region. Since this region is in the past of the $t>t_h$ region, the interior observer perceives it as a white hole.

We notice here the important aspect that the line element in Eq.(5) describes the region that classically refers to the exterior region ($r>r_h$) of the black hole spacetime as being a gravitationally trapped region ($0<t<t_h$) in the case where the characteristics of space and time are switched. In addition, we should remark that $t$ is the timelike coordinate in this region of spacetime for both an external observer and an internal observer.    

We should specify for the sake of completeness that for $t>t_h$, in the classical interior black hole region we have a line element of the form:

\begin{equation}
ds^2=\frac{\ell^2}{t^2-t^{2}_{h}}dt^2-\frac{t^2-t^{2}_{h}}{\ell^2}dr^2+t^2d\phi^2
\end{equation}

which describes a region that extends to infinity as perceived by an interior observer.

\section{The maximal extension of the BTZ spacetime}

We would like to extend the interior line element to better understand the correlations with the exterior solution. Following the usual prescription that can be seen explicitly in APPENDIX B we arrive at the metric for the maximally extended BTZ spacetime, in terms of $U,V$ coordinates, of the form:

\begin{equation}
ds^2=\frac{4\ell^2}{ \left(1+UV \right)^2} \left[-dUdV+\frac{t^{2}_{h}}{4\ell^2} \left(1-UV \right)^2 d\phi^2 \right]
 \end{equation}

with $U<0$ and $V>0$.

Notably, we recovered a relation similar to the exterior solution. The notable difference is that the two metrics (interior and exterior) refer to different regions of spacetime.

The original BTZ coordinates $(r,t)$ are related to the coordinates $(U,V)$ by the relations:  

\begin{equation}
\frac{t}{t_h}=\frac{1-UV}{1+UV}
\end{equation}

and
\begin{equation}
r=\frac{\ell^2}{2t_h} ln \left(\frac{V}{-U} \right)
\end{equation}

We can easily probe from eq.(8) that $t=t_h$ corresponds to $UV=0$, i.e. either $U=0$ or $V=0$ at the event horizon $t_h$. Additionally, the singularity $t=0$ corresponds to the right branch of the hyperbolae, $UV=1$, since now the region I is described by the null coordinates $U<0$ , $V<0$ as is illustrated in the Fig.1. This result moves the singularity from region $II$ to the region $I$ while the region $II$ becomes the exterior region described by the null coordinates, $U<0$ , $V>0$. In other words, the $t \rightarrow \infty$ limit corresponds to the upper branch of the hyperbolae, $UV=-1$.

\begin{figure}
\includegraphics[width=8.6cm]{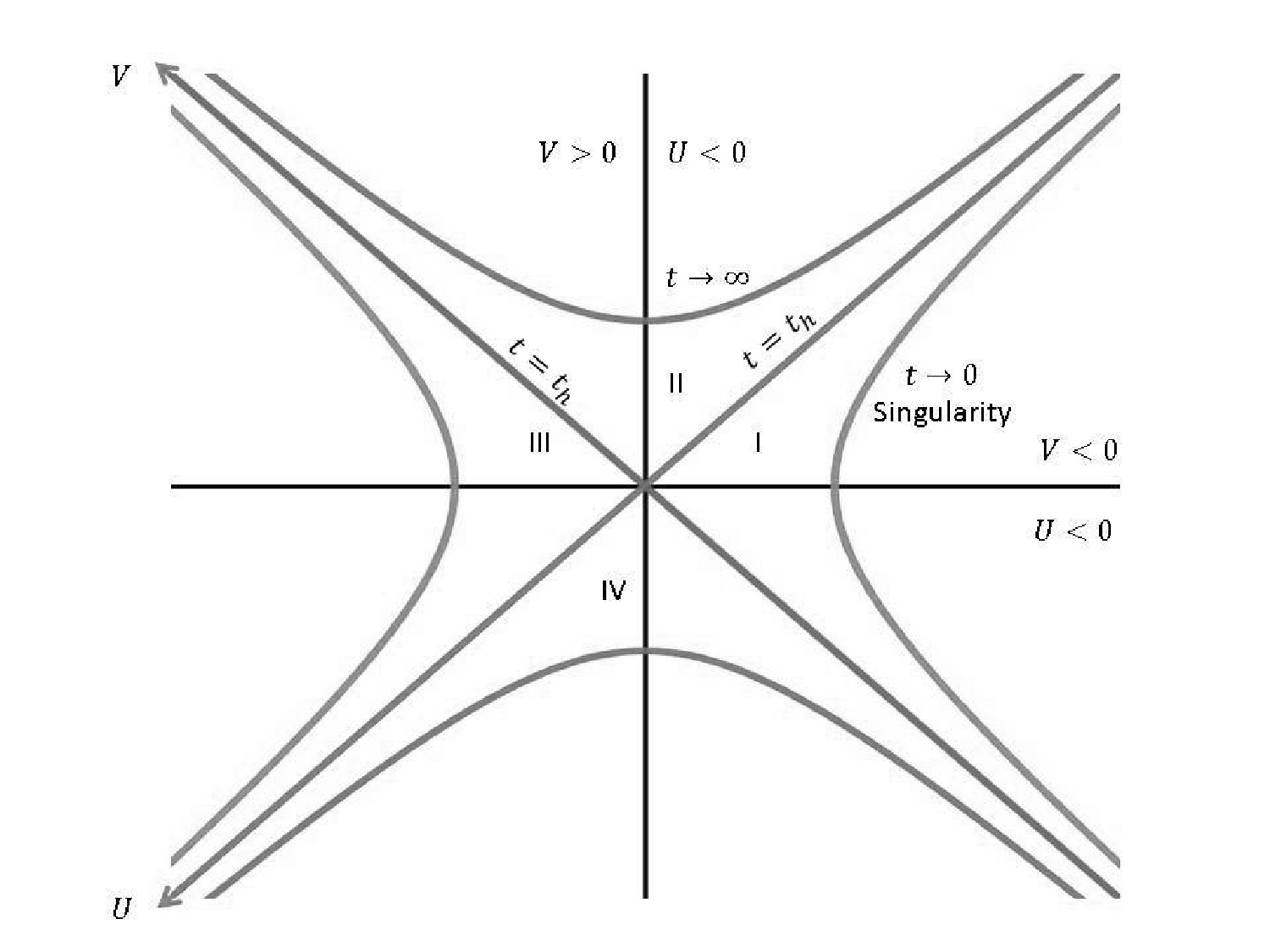}
\caption{\label{fig:fig1} Spacetime diagram of the maximally extended BTZ black hole.}
\end{figure}

To conclude, we have arrived at the spacetime shown in Fig.1. by starting with a metric describing only region II $(U < 0,V > 0)$, which classically refers to the  interior of the black hole with $t > t_h$ . However, we further extend the spacetime through the event horizon at $t= t_h$ into regions with $t < t_h$ , namely the white hole and black hole regions, $I$  $(U<0 ,V< 0)$ and $III$ $(U>0 ,V> 0)$, respectively . Both of these regions contain true singularities such as $t \rightarrow 0$. In addition, examining the metric in Eq.(6) we can  show that $t$ is spacelike and $r$ is timelike in the regions $II$ and $IV$ while $t$ is timelike and $r$ is spacelike in the regions $I$ and $III$. This result is similar to that of the exterior region. 

\section{Euclidean metric of the interior BTZ solution}

We conclude the part that refers to the BTZ interior solution by briefly considering the Euclidean metric. Thus, let us relate once again to the metric behind the horizon given in Eq.(6) and consider the simple case with $t_h=\ell=1$ such that the metric is now:

\begin{equation}
ds^2=\frac{1}{ t^2 -1}dt^2 - \left(t^2 -1 \right)dr^2+t^2 d\phi^2 
 \end{equation}
We consider here a Wick rotation with $\tilde r=ir$  since now the $r$ coordinate is timelike. The Euclidean metric of the BTZ spacetime takes the form:  

\begin{equation}
ds^2=\frac{1}{t^2 -1}dt^2 + \left(t^2 -1 \right)d\tilde r^2+t^2 d\phi^2 
 \end{equation}

On the other hand, with the $t_h=\ell=1$ simplification, the metric of the BTZ spacetime in coordinates $(U,V)$ is of the form:
\begin{equation}
ds^2=\frac{4}{ \left(1+UV \right)^2} \left[-dUdV+ \frac{1}{4}\left(1-UV \right)^2 d\phi^2 \right]
 \end{equation}

We note here that both eq.(11) and eq.(12) are similar to their exterior counterparts. As such is easy to determine that they are equivalent through the analytical continuation in the imaginary part with $z=-u$ and $\bar z=v$, where $z=|z|e^{i\tilde r}$ and $r=r+\beta$, $\phi=\phi+2 \pi$.

\begin{equation}
ds^2=\frac{4}{\left(1-|z|^2 \right)^2} \left[dzd\bar z+\frac{1}{4} \left(1+|z|^2 \right)^2 d\phi^2 \right]
 \end{equation}

Once again the Euclidian metric can be glued to the spatial section which is now a time-like section as depicted in Fig.2.

\begin{figure}
\includegraphics[width=6.6cm]{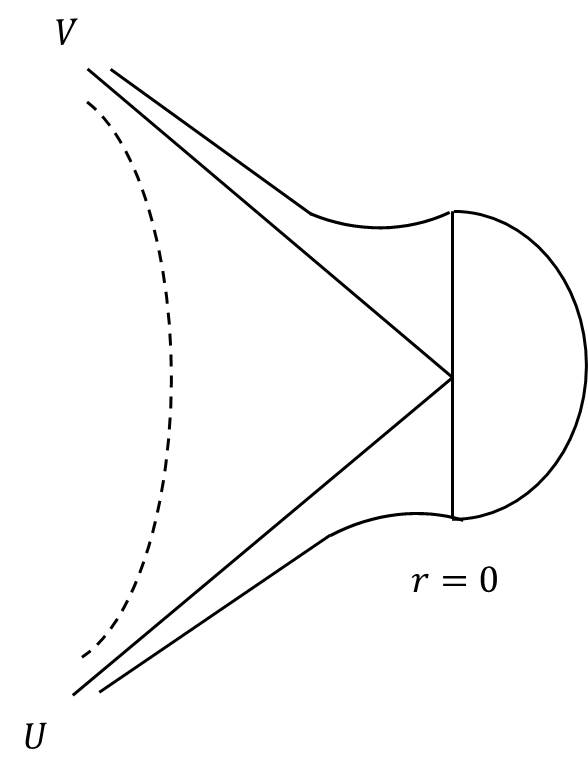}
\caption{\label{fig:fig2} Gluing half Euclidean geometry to half Lorentzian geometry at r=0.}
\end{figure}

We can also infer from the  fig.2 that the Euclidean part of the geometry gives the initial wave function which is then evolved in Lorentzian signature \cite{mal} as we will argue in the next section.

\section{The thermofield double state dual to the interior solution}

Since an observer behind the BTZ black hole horizon probes exactly the same spacetime geometry as the exterior observer does, we should expect a similar AdS/CFT correspondence even when the space and time roles of the BTZ metric are interchanged. In other words, we should be able to find a thermofield double state as dual to the BTZ interior solution.

Although the derivation of the thermofield double state dual to the BTZ interior metric exactly mirrors its classical counterpart we consider here a brief explanation just to make the connection to the last part of the paper. 
In this respect, since now $r$ coordinate is time-like we would consider evaluating the path integral over an angle $\pi$ along the angular direction $r$. That is, to express the vacuum (ground) state $\ket{\bar 0}$  in the basis of eigenstates we take the left $\bar L$ and right $\bar R$ wedges along the $r$ coordinates, as schematically shown in fig.3.

\begin{figure}
\includegraphics[width=9.6cm]{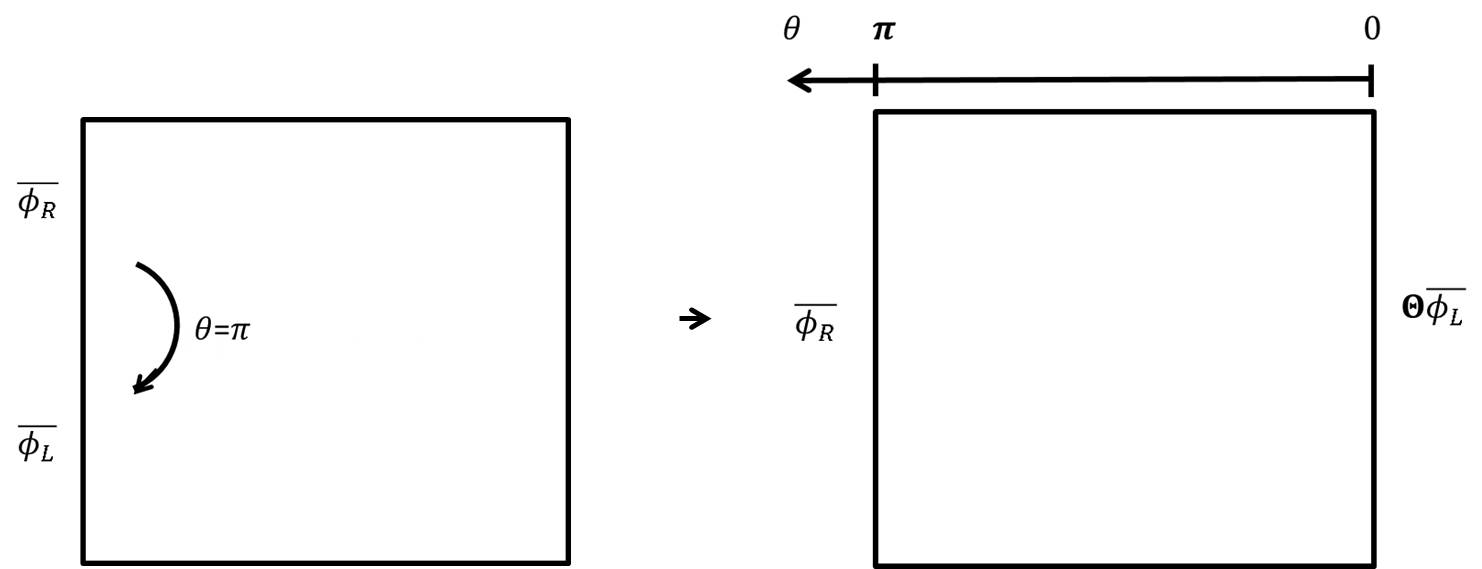}
\caption{\label{fig:fig3} Vacuum (ground) state dual to the BTZ interior solution.}
\end{figure}

As a result, the boundary two quantum systems, $\bar L$  and $\bar R$ , defined on the Hilbert space $\bar{\mathcal H}={\bar{\mathcal H}_{\bar L}} \otimes {\bar{\mathcal {H}}_{\bar R}}$, when evaluated in the Euclidean path integral formalism yields:

\begin{equation}
\bra{\bar{\phi}_{\bar L}\bar{\phi}_{\bar R}}\ket{\bar 0} =\bra{\bar{\phi}_{\bar R}} e^{-{\pi} \bar H} \Theta \ket{\bar{\phi}_{\bar L}}
\end{equation}

Here, the Hamiltonian $\bar{H}$ generates rotations by an angle $\pi$ in the Euclidean plane with $r$ coordinate as time-like. Furthermore, we evaluate the expression in Eq.(14) as usual by inserting a complete set of eigenstates of the Hamiltonian $\bar{H}$ , \cite{har}:

\begin{equation}
\begin{split}
\bra{\bar{\phi}_{\bar L}\bar{\phi}_{\bar R}}\ket{\bar 0}=\sum_{i} e^{-{\pi}  \bar{E}_{\bar i}} \bra{\bar i} \Theta \ket{\bar{\phi}_{\bar L}} \bra{\bar{\phi}_{\bar R}} \ket{\bar i}\\
=\sum_{i} e^{-{\pi}  \bar{E}_{\bar i}} \bra{\bar{\phi}_{\bar L}} \ket{\bar {i^*}} \bra{\bar{\phi}_{\bar R}} \ket{\bar i}
\end{split}
\end{equation}

Thus, we can write the vacuum state expression as:
\begin{equation}
\ket{\bar 0} =\sum_{i} e^{-{\pi}  \bar{E}_{\bar i}} \ket{\bar {i^*}}  \ket{\bar i}
\end{equation}

Now, considering the inverse temperature $\beta = 2 \pi$ and from the normalization $\bra{\bar 0} \ket{\bar 0} = 1$ we obtain 

\begin{equation}
\ket{\overline{TFD}}=\frac{1}{\sqrt{Z}} \sum_i e^\frac{-\beta \bar{E}_{\bar i}}{2} \ket{\bar i^*}  \ket{\bar i}
\end{equation}

which is the thermofield double state associated with the interior solution. Indeed, we should note that it is similar to its counterpart dual to the exterior BTZ solution, as expected.

\section{Non-orientable spacetime of the eternal black hole} 

Let us now imagine a thought experiment in which we attempt to clarify the particularities of switching the space and time roles behind the horizon in relation to AdS/CFT correspondence. Thus, we consider an (hypothetical) observer initially located in region $I$, i.e. the exterior region of spacetime. Here, the line element of the BTZ spacetime is:

\begin{equation}
ds^2=-\left(r^2 -1 \right)dt^2 + \frac{1}{r^2 -1}dr^2+r^2 d\phi^2
 \end{equation}

since in region $I$ the coordinate $r$ is space-like and the coordinate $t$ is time-like. In Kruskal coordinates $(U,V)$ the BTZ metric is:

\begin{equation}
ds^2=\frac{4}{ \left(1+UV \right)^2} \left[-dUdV+ \frac{1}{4}\left(1-UV \right)^2 d\phi^2 \right]
 \end{equation}

which is dual to the thermofield double state \cite{mal}, \cite{suss}. As emphasized in the previous section, the thermofield double state can be derived via the Euclidean path integral formulation (considering the $t$ coordinate as time-like) as: 

\begin{equation}
\bra{\phi_L \phi_R}\ket{ 0} =\bra{\phi_R} e^{-\pi H} \Theta \ket{\phi_L}
\end{equation}

Now, let our infalling observer that travels on an ingoing radial geodesic cross the event horizon of the black hole. Behind the horizon in region $II$ the roles of space and time are switched such that the coordinate $t$ is space-like and the coordinate $r$ is time-like. Nevertheless, we have a similar BTZ metric as in eq. (18), this time with the coordinates $r$ as time-like  

\begin{equation}
ds^2=\frac{1}{ t^2 -1}dt^2 - \left(t^2 -1 \right)dr^2+t^2 d\phi^2 
 \end{equation}

while in Kruskal $U,V$ coordinates the BTZ metric remains unchanged.  

As we have seen previously, behind the horizon, the Euclidean path integral formulation we should consider is expressed as:

\begin{equation}
\bra{\bar{\phi}_{\bar L}\bar{\phi}_{\bar R}}\ket{\bar 0} =\bra{\bar{\phi}_{\bar R}} e^{-{\pi} \bar H} \Theta \ket{\bar{\phi}_{\bar L}}
\end{equation}

The thermofield double state that results from evaluating this expression connects in the gravity dual the region $II$ and the region $IV$ of the BTZ black hole diagram. Accordingly, once in region $II$ of spacetime, the observer notice a black hole that can be understood as the thermal state which results from tracing over the degrees of freedom of the left quantum theory:    

\begin{equation}
\bar{\rho}_{\bar R} = Tr_{\bar L} \bra{\overline{TFD}} \ket{\overline{TFD}} =\frac{1}{Z} \sum_i e^{-{\beta} \bar{E}_{\bar i}} \ket{\bar i} \bra{\bar i}
\end{equation}

Finally, our observer can travel once again on ingoing radial geodesics toward the black hole. Once the observer crosses the black hole’s event horizon it reaches the region $III$ where the space and time switch their roles again.  

We would like to synthesize the journey of our observer throughout the black hole spacetime in a unitary and consistent way in the path integral formalism. Thus, in the above scenario, to construct the Euclidean path integral we have to consider the full Hilbert space $\mathcal H_{Total}=\mathcal H \otimes \bar {\mathcal H}$  with $\mathcal H= \mathcal{H}_L \otimes \mathcal{H}_R$ and, as we have seen, $\bar{\mathcal H}= {\bar{\mathcal {H}}_{\bar L}} \otimes  {\bar{\mathcal {H}}_{\bar R}}$. 

Indeed, we can decompose the total Hilbert as $\mathcal H_{Total}= \left(\mathcal{H}_L \otimes {\bar{\mathcal {H}}_{\bar L}}\right) \otimes \left(\mathcal{H}_R \otimes {\bar{\mathcal {H}}_{\bar R}}\right)$. Now, considering $\mathcal H_{LS}=\mathcal{H}_L \otimes {\bar{\mathcal {H}}_{\bar L}}$ as the left sector (the white sector in Fig.4 ) and $\mathcal H_{RS}=\mathcal{H}_R \otimes {\bar{\mathcal {H}}_{\bar R}}$   as the right sector (the gray sector in Fig.4) we can define the total Hilbert space $\mathcal H= \mathcal{H}_{LS} \otimes \mathcal{H}_{RS}$ . 

\begin{figure}
\includegraphics[width=7.6cm]{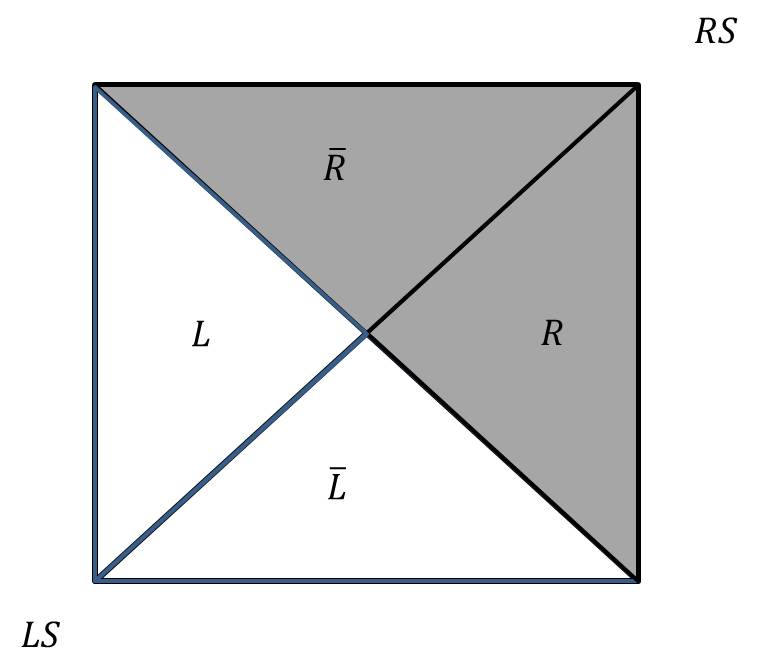}
\caption{\label{fig:fig4} Left and right sectors of the full BTZ black hole.}
\end{figure}

From the path integral perspective we can argue that in order to cover the entire black hole spacetime we should consider both vacuum states:   

\begin{equation}
\bra{\phi_L \phi_R \bar{\phi}_{\bar L}\bar{\phi}_{\bar R}}\ket{0 \bar 0} =\bra{\phi_R \bar{\phi}_{\bar R}} e^{-{\pi}H} e^{-{\pi} \bar H} \ket{\phi_L \bar{\phi}_{\bar L}}
\end{equation}

as it can intuitively be seen in Fig.5.

\begin{figure}
\includegraphics[width=9.6cm]{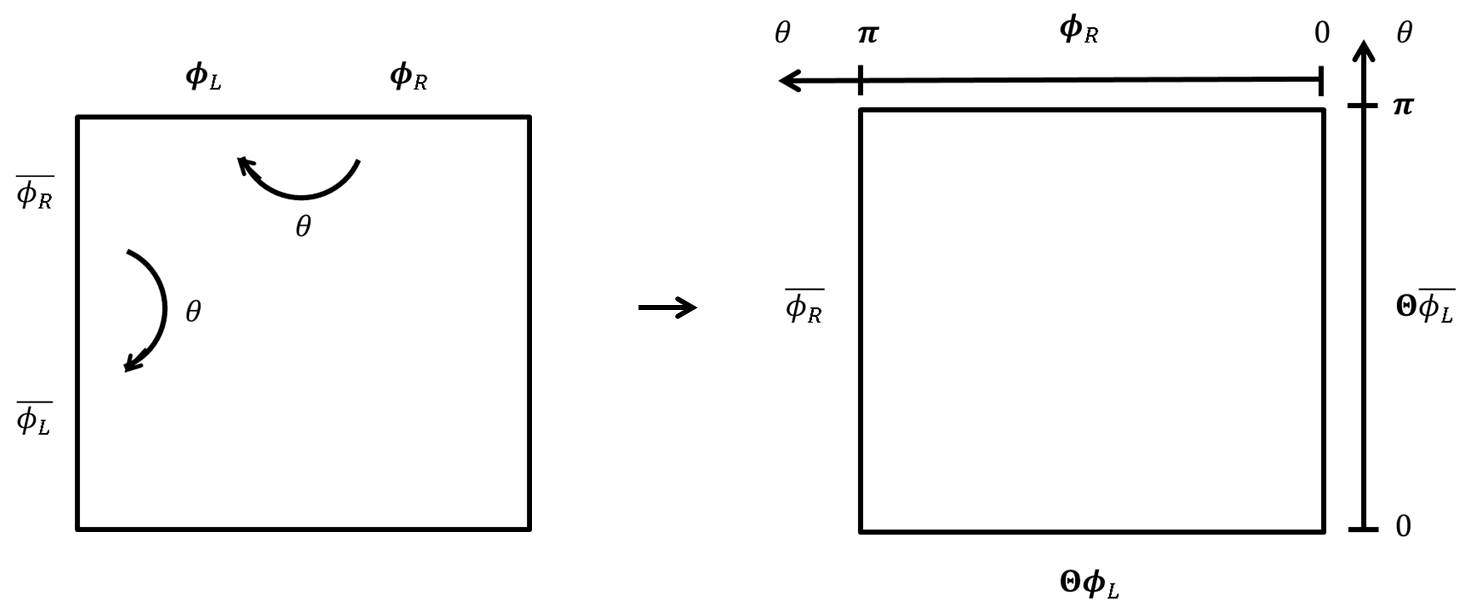}
\caption{\label{fig:fig5} Euclidean path integral representation of the vacuum (ground) state.}
\end{figure}

We can notice from fig.5  that the Hamiltonian $H$ generates rotation by $\pi$ in the Euclidean plane at $t=0$, whereas the Hamiltonian $\bar H$ generates rotation by $\pi$  in the Euclidean plane at  $r=0$. The total Hamiltonian that generates both rotations over an angle $\pi$  , one in the Euclidean plane with $t$ as the time-like coordinate and one in the Euclidean plan having $r$ as time-like coordinate,  would be $ H_{Total}=H+\bar H$.

With all these clarifications in mind, let us now evaluate the Eq.(24) by inserting a complete set of eigenstates for $H$ and $\bar H$ to obtain:

\begin{equation}
\begin{split}
\bra{\phi_L \phi_R \bar{\phi}_{\bar L}\bar{\phi}_{\bar R}}\ket{0 \bar 0}=\sum_{i} e^{-{\pi}  E_{i}} e^{-{\pi} \bar{E}_{\bar i}} \bra{i \bar i} \Theta \ket{\phi_L \bar{\phi}_{\bar L}} \bra{\phi_R \bar{\phi}_{\bar R}} \ket{i \bar i}\\
=\sum_{i} e^{-{\pi}  E_{i}} e^{-{\pi} \bar{E}_{\bar i}} \bra{\phi_L \bar{\phi}_{\bar L}}  \ket{i^* \bar {i^*}} \bra{\phi_R \bar{\phi}_{\bar R}} \ket{i \bar i}
\end{split}
\end{equation}

Thus, we arrive at the simple expression: 

\begin{equation}
\ket{0 \bar0}=\sum_{i} e^{-{\pi}  E_{i}} e^{-{\pi} \bar{E}_{\bar i}} \ket{i^* \bar {i^*}} \ket{i \bar i}
\end{equation}

Rearranging the terms conveniently in the above expression we remain with 
\begin{equation}
\ket{0 \bar{0}}=\sum_{i} \bra{i \bar i} e^{-{\pi}  E_{i}} e^{-{\pi} \bar{E}_{\bar i}} \ket{i \bar i}
\end{equation}

Now, from normalization $\bra{0 \bar{0}} \ket{0 \bar{0}}=1$ we can derive      

\begin{equation}
Z=\sum_{i} \bra{i \bar i} e^{-{\pi}  E_{i}} e^{-{\pi} \bar{E}_{\bar i}} \ket{i \bar i}
\end{equation}

In this path integral we may recognize the partition function of the Klein bottle geometry \cite{berr}, \cite{ryu}, \cite{wei}, \cite{hao}. Consequently, for the full BTZ black hole spacetime resulting from the Kruskal diagram in Fig.1, if we consider regions $I$ and $II$  as the right segment and regions $III$ and $IV$ as the left segment, the spacetime of the black hole bulk should have a non-orientable spacetime. 

We like to briefly mention here that a similar result is extensively used in the study of many-body topological phases of matter \cite{shi}, \cite{shap}, \cite{cho}, \cite{poll}.   Specifically, we refer to topological invariants of many-body symmetry-protected topological (SPT) phases protected by orientation reversing symmetry. Within the Euclidean path integral, orientation reversing symmetry (such as time-reversal) can be used to twist boundary conditions, leading to non-orientable spacetime such as the real projective plane, Klein bottle, etc.

Let us return to the normalized eq.(26) and take the inverse temperature $\beta=2 \pi$ such that we can find the thermofield double-like state 

\begin{equation}
\ket{TFD_{total}}=\frac{1}{Z} \sum_i e^\frac{-\beta (E_i + {\bar{E}}_{\bar i})}{2} \ket{i \bar i} \Theta \ket{i \bar i}
\end{equation}

which is dual to the non-orientable spacetime of the full eternal black hole geometry. 

\section{Conclusions}

In this paper, we emphasized over the importance of considering the interchange of space and time roles in the context of the BTZ solution. Accordingly, we have derived the BTZ metric for the special case of switching space and time roles which can be considered an internal line element. We show that the maximally extended geometry of this interior solution is similar to its exterior solution counterpart and has the same Euclidean metric. Consequently, we construct the thermofield double state dual to this special internal solution.

In this scenario, when we consider both thermofield double states as dual to the full BTZ black hole, we evaluate the partition function of the bulk. The partition function reveals a non-orientable spacetime.

We conclude by deriving the thermofield double-like state that connects regions of spacetime with opposite orientations of space and time in the gravity dual.

\renewcommand{\theequation}{A-\arabic{equation}}
  % redefine the command that creates the equation no.
  \setcounter{equation}{0}  % reset counter 
  \section*{APPENDIX A: Derivation of the interior BTZ metric }

To derive the interior BTZ metric we start by considering a general radially symmetric line element:

\begin{equation}
ds^2=-B^2(t)dt^2+A^2(t)dr^2+F^2(t)d\phi^2.
\end{equation}

Then, taking into account the spherical symmetry, and since we assumed that t is the radial coordinate we rewrite the metric as:

\begin{equation}
ds^2=-B^2(t)dt^2+A^2(t)dr^2+t^2d\phi^2
\end{equation}

and we attempt to derive the functions  $B(t)$ and $A(t)$. 
The metric should describe the spacetime as seen by an observer situated behind the event horizon such that it would satisfy the Einstein’s equations. Thus, we can write the vielbeins: 

\begin{equation}
e^0=B(t)dt,\     \  e^1=A(t)dr,\      \   e^2=td\phi.
\end{equation}

The components of the spin connection can be found using the Cartan’s first structure equation $de^{\mu}=-\omega^{\mu}_{\nu}\wedge e^{\mu}$  as:

\begin{equation}
de^0=B^{'}(t)dt\wedge dt=0,
\end{equation}

since $dt\wedge dt=0$.

Furthermore, we identify

\begin{equation}
de^1=A^{'}(t)dt\wedge dr,
\end{equation}

and

\begin{equation}
de^2= dt\wedge d\phi.
\end{equation}

Now, we may infer that the only term with $dt \wedge dr$ is $\omega^{2}_{r} dr \wedge e^0$, such that we have: 

\begin{equation}
A^{'}(t)dt\wedge dr=-\omega^{2}_{r} dr \wedge B(t)dt,
\end{equation}

Thus, we can write 

\begin{equation}
\omega^{2}_{r} = -\frac{ A^{'}(t)}{ B(t)},
\end{equation}

Similarly, observing that the only term with $dt \wedge d\phi$ is $\omega^{1}_{\phi} d\phi \wedge e^0$ we conclude that 

\begin{equation}
dt\wedge d\phi=\omega^{1}_{\phi} d\phi \wedge B(t)dt,
\end{equation}

and further 

\begin{equation}
\omega^{1}_{\phi} = -\frac{1}{B(t)}.
\end{equation}

All the other terms vanished.

Thus, to summarize, we have

\begin{equation}
\omega^0=0,\     \  \omega^1=-\frac{1}{ B(t)},\      \   \omega^2=-\frac{ A^{'}(t)}{ B(t)}.
\end{equation}

Now we will find the curvature two-form  $R^{\mu}=d\omega^{mu}+\frac{1}{2}\omega^{\mu}_{\nu} \wedge \omega^{nu}$  and solve the equation for the empty space  $R^{\mu}=\frac{\Lambda}{2} e^{\mu}_{\nu} \wedge e^{\nu}$  with the cosmological constant $\Lambda$. 

To find $A(t)$ and $B(t)$ we need only two of the three equations that result:

\begin{equation}
R^0 = -\frac{A^{'}(t)}{ B^2(t)} dr \wedge d\phi=\Lambda A(t) t dr \wedge d\phi.
\end{equation}

and
\begin{equation}
R^1 = \frac{B^{'}(t)}{ B^2(t)} dt \wedge d\phi=-\Lambda B(t) t dt \wedge d\phi.
\end{equation}

since the equation for $R^1$ is only in terms of $B(t)$. 
We can form the differential equations

\begin{equation}
\begin{split}
\frac{A^{'}(t)}{B^2(t)}=-\Lambda A(t)t \\
\frac{B^{'}(t)}{B^2(t)}=-\Lambda B(t)t 
\end{split}
\end{equation}

Let us integrate the last differential equation with the separation of variables: 

\begin{equation}
\frac{B^{'}(t)}{ B^3(t)} =-\Lambda t.
\end{equation}

which yields:

\begin{equation}
-\frac{1}{ B^2(t)} =-\left(\Lambda t^2+M \right).
\end{equation}

where we set $M$ as the constant of integration. Thus we have:

\begin{equation}
B^2(t) =\frac{1}{\Lambda t^2+M} .
\end{equation}
 
If we take the AdS spacetime cosmological constant $\Lambda =-\frac{1}{\ell^2}$ we remain with:

\begin{equation}
B^2(t) =\frac{1}{M-\frac{t^2}{\ell^2}} .
\end{equation}

Now replacing $B^2 (t)$ in eq.(A-14) we find

\begin{equation}
\frac{A^{'}(t)}{A(t)}=\frac{-\Lambda t}{\Lambda t^2+M } .
\end{equation}

which by integration results in: 

\begin{equation}
A^{2}(t)=\Lambda t^2+M,
\end{equation}

or substituting the cosmological constant $\Lambda$ yields:

\begin{equation}
A^{2}(t)=M-\frac{t^2}{\ell^2}.
\end{equation}

With the help of eq.(A-18) and eq.(A-21) we can write the full metric of the 2+1 black hole

\begin{equation}
ds^2=-\frac{1}{\left(M-\frac{t^2}{\ell^2}\right)}dt^2+\left(M-\frac{t^2}{\ell^2}\right)dr^2+t^2d\phi^2
\end{equation}

which is only valid for values  $\frac{t^2}{\ell^2}<M$. As such, the line element describes the region that refers to the exterior of the black hole spacetime. However, the metric in Eq.(A-22)  describes the region $I$ as being a gravitationally trapped region in stark contrast with the exterior BTZ solution. We may remark that interchanging the roles of space and time the exterior region is mapped to a trapped region. Additionally, we note that in this region of spacetime, $t$ is the timelike coordinate for both an external observer and an internal observer.

In the interior BTZ black hole region we have a line element of the form:

\begin{equation}
ds^2=-\frac{1}{\left(\frac{t^2}{\ell^2}-M\right)}dt^2+\left(\frac{t^2}{\ell^2}-M\right)dr^2+t^2d\phi^2
\end{equation}

which describes a region that extends to infinity as perceived by an interior observer.
Let us consider the metric behind the horizon in eq.(A-23) , rearrange the terms for convenience and take $M \ell^2=t^{2}_h$ such that we find the line element:

\begin{equation}
ds^2=-\frac{\ell^2}{t^2-t^{2}_{h}}dt^2+\frac{t^2-t^{2}_{h}}{\ell^2}dr^2+t^2d\phi^2
\end{equation}

Here $t \rightarrow t_h$ is the event horizon while we still have a singularity at $t \rightarrow 0$ in the region with $t<t_h$ which is the classical exterior region. Since this region is in the past of the $t>t_h$ region the interior observer perceives it as a white hole.

\renewcommand{\theequation}{B-\arabic{equation}}
  % redefine the command that creates the equation no.
  \setcounter{equation}{0}  % reset counter 
  \section*{APPENDIX B: The maximal extension of interior BTZ metric }

We start with the metric 

\begin{equation}
ds^2=-\frac{\ell^2}{t^2-t^{2}_{h}}dt^2+\frac{t^2-t^{2}_{h}}{\ell^2}dr^2+t^2d\phi^2
\end{equation}

and we look for new coordinates related to null geodesics with constant $\phi$ that in our case must fulfil

\begin{equation}
dr=\pm\frac{\ell^2}{t^2-t^{2}_{h}}dt
\end{equation}

That is, since $t$ is the spacelike coordinate we define the new coordinate $t_{*}$ 

\begin{equation}
dt_{*}=\frac{\ell^2}{t^2-t^{2}_{h}}dt
\end{equation}

such that the null geodesics with constant $\phi$ are now given by $dr=\pm dt_{*}$ or
\begin{equation}
r \mp t_{*}=const
 \end{equation}

We solve the differential equation in eq.(B-3) under the condition of a zero constant of integration such that $t_{*} \rightarrow 0$ corresponds to $t \rightarrow \infty$, to find:

\begin{equation}
t_{*}=-\frac{\ell^2}{t_h} arccoth \left(\frac{t}{t_h} \right)
 \end{equation}

since the line element in Eq.(B-1) describes the region with $t>t_h$ (the interior region ). From Eq. (B-5)  we deduce further

\begin{equation}
t=-t_h \coth \left(\frac{t_h t_*}{\ell^2} \right)
 \end{equation}

With this new coordinate in hand the metric takes the form:

\begin{equation}
ds^2=\frac{t^{2}_{h}}{\ell^2 \sinh^2\left(\frac{t_h t_*}{\ell^2} \right)} \left[dt^{2}_{*}-dr^2+\ell^2 \cosh^2\left(\frac{t_h t_*}{\ell^2} \right)]d\phi^2 \right]
 \end{equation}

Now, returning to eq.(B-4), the two classes of null geodesics are described by $u=const$ and  $v=const$ with:

\begin{equation}
u=r-t_{*},\      \  v=r+t_{*}
 \end{equation}

Using these new coordinates $u,v$ we can now calculate the metric as:

\begin{equation}
ds^2=\frac{t^{2}_{h}}{\ell^2 \sinh^2\left(\frac{t_h (v-u)}{2\ell^2} \right)} \left[-dudv+\ell^2 \cosh^2\left(\frac{t_h (v-u)}{2\ell^2} \right)]d\phi^2 \right]
 \end{equation}

We finally arrive at the point where we have to compactify the coordinates. First thing to notice here is that now the interior BTZ region is the region that extends to infinity such that we should introduce the coordinates $U<0$ and $V>0$ with  $U=-e^-\frac{t_{h}u}{\ell^2}$ and  $V=e^\frac{t_{h}v}{\ell^2}$  . To complete the picture of the new coordinates $U,V$ we emphasize here that in the classical exterior region we now have $U<0$ and $V<0$.
 The metric for the maximal extended BTZ spacetime in terms of $U,V$ coordinates takes the form:
\begin{equation}
ds^2=\frac{4\ell^2}{ \left(1+UV \right)^2} \left[-dUdV+\frac{t^{2}_{h}}{4\ell^2} \left(1-UV \right)^2 d\phi^2 \right]
 \end{equation}

with $U<0$ and $V>0$.
It can easily be noted here that we recover a relation similar to the exterior solution. The notable difference is that the two metrics (interior and exterior, respectively) refer to different regions of  spacetime.

\end{document}